\documentstyle[12pt]{article}
\begin{document}

\newcommand{\bra}[1]{\langle #1 \vert}
\newcommand{\ket}[1]{\vert #1 \rangle}
\newcommand{\beq}{\begin{equation}}
\newcommand{\eeq}{\end{equation}}
\newcommand{\avet}[1]{\langle\!\langle {#1} \rangle\!\rangle}
\def\le#1{\label{eq:#1}}
\def\re#1{\ref{eq:#1}}
\def\barr{\begin{array}[b]}
\def\barc{\begin{array}}
\def\bart{\begin{array}[t]}
\def\ear{\end{array}}

%\draft

\noindent
\centerline{\bf Low densities in nuclear and neutron matters and in the nuclear 
surface}
\vskip 1 cm
\centerline{\it M. Baldo$^{1}$, C. Maieron$^{1}$, 
P. Schuck$^{2}$ and X. Vi\~nas$^3$}
\vskip 0.3 cm
\centerline{$^{1}$
Istituto Nazionale di Fisica Nucleare, Sezione di Catania}
\centerline{Via Santa Sofia 64, I-95123 Catania, Italy}
\vskip 0.3 cm
\centerline{$^2$ Institut de Physique Nucleaire IN2P3 - CNRS }  
\centerline{Universit\'e Paris-Sud, 91406  Orsay C\`edex, France}
\vskip 0.3 cm
\centerline{$^3$Departament d'Estructura i Constituents de la Mat\`eria,}
\centerline{ Universitat de Barcelona,
 Av. Diagonal 647, E-08028 Barcelona, Spain}

\newpage
\vskip 3 cm
\noindent
{\bf Abstract.}
Nuclear and neutron matters are investigated in the low density region,
well below the nuclear saturation density.
Microscopic calculations, based on the Bethe-Brueckner approach
with a few realistic nucleon-nucleon potentials,
are compared with the predictions of a set of phenomenological 
effective interactions, mostly employed in nuclear structure studies.
An energy functional is constructed on the basis of the microscopic 
bulk EoS and applied to a selection of nuclei throughout the mass table.
The results provide a microscopic basis for a link between 
nuclear surface behaviour and neutron EoS previously observed
with phenomenological effective forces.
Possible effects of pairing on asymmetric nuclear matter 
are also analyzed in detail.
The results are expected to illuminate the physical mechanisms
which determine the behaviour of the surface density tail in exotic nuclei.

%\pacs{PACS:
%21.65.+f,  % Nuclear matter
%24.10.Cn}  % Many-body theory

\newpage

\par\noindent
{\bf Introduction}
\par
Most of the studies of symmetric and asymmetric nuclear matter 
have been restricted to densities from about saturation ($\sim 0.15$ fm$^{-3}$)
to few times this value. This density interval is the most
important for the physics of neutron stars and of heavy ion collisions,
both at intermediate and relativistic energies.\par
In finite nuclei, the surface region is characterized by a density profile
the detailed shape of
which is determined in nuclear structure calculations
by the gradient term in phenomenological effective
nucleon-nucleon (NN) forces such as the Skyrme forces \cite{sly4,skm},
or by a finite range interaction as in the case with the Gogny
force \cite{gog}.
In exotic nuclei, with large asymmetry,
the density drops at the surface with a relatively smooth profile.
The neutron to proton ratio can be quite different from the bulk one,
which gives rise to phenomena like the appearance of a 
neutron skin or of a neutron halo. But even for normal nuclei the asymmetry 
in the far tail may vary substantially from the one in the bulk.
In a Local Density Approximation (LDA) picture the  
detailed asymmetry structure, and thus also the surface profile, 
should be related to the properties of asymmetric nuclear matter
at density much lower than the saturation one. 
In particular, the symmetry energy $a_y$ at low densities,
which is embodied in the effective force, should play a major role in
fixing the neutron to proton ratio throughout the surface region.
These aspects may also be of importance in other low density
nuclear systems such as the expansion phase after central heavy ion
collisions at intermediate energies or during the collapse of massive
stars.\par
It therefore appears of great interest to investigate 
in detail the asymmetry properties of low density nuclear matter.
This shall be done in this work using firstly a microscopic
Brueckner Hartree-Fock approach starting from the bare NN force
and secondly with current phenomenological forces of the Skyrme
and Gogny type. The similarities and differences between both
approaches shall be analyzed. \par
A semi-empirical connection between low density nuclear matter 
Equation of State (EoS) and finite nuclei surface properties has been
presented in ref. \cite{brown}. The difference between neutron
and proton root mean square radii in $^{208}Pb$ was calculated   
for an extended set of Skyrme interactions. This
difference $ r_n \, -\, r_p $ turns out to be 
linearly correlated with the slope
$\chi$ of the corresponding neutron matter EoS at the chosen density
$\rho = 0.1$ fm$^{-3}$. This slope, i.e. the derivative of the energy 
per particle with respect to density, can largely differ from one Skyrme
force to another, since no direct phenomenological constraint exists for 
neutron matter. This study has been extended to relativistic mean field
functionals in ref. \cite{brown2}, and the same linear correlation
has been found in ref. \cite{xavier}, where the analysis
comprises a large set of energy functionals. 
A partial explanation of the correlation was presented
in ref. \cite{Dieper} on the basis of Landau Fermi liquid theory. 
However, this linear correlation still appears surprising, since 
it is well known that in order to properly
describe the surface properties of finite nuclei the bulk term of the
energy functional must be implemented with gradient terms, 
not included in ref \cite{Dieper}, which should also
play a role in determining the value of $ r_n \, -\, r_p $. 
One of the main aims 
of the paper is to establish to what extent the bulk part of the
functional, which is directly related to nuclear matter EoS,
dominates indeed the nuclear surface behaviour. To this purpose,
we shall construct an energy functional based on many-body calculations,
in particular for neutron matter, for which the EoS seems to be well 
established on a microscopical basis. This will be used to restrict 
the range of possible values of $ r_n \, -\, r_p $. 
\par
A further important aspect of low density asymmetric
nuclear matter may be its pairing properties. Indeed we will find
that low density nuclear matter with asymmetry such that it is
slightly unbound can become bound solely through the action of
neutron-neutron pairing which is quite strong at low densities.
This phenomenon may have important consequences in the formation
of neutron-skins and -haloes in very neutron rich nuclei.\par
In detail the paper is organized as follows.
In Sec. 2 we give a brief introduction to the microscopic many-body
theory and present the results for different 
nucleon-nucleon realistic interactions, for symmetric nuclear matter,
pure neutron matter and for asymmetric nuclear matter in general.
In Sec. 3 the results for a few phenomenological NN effective forces
are presented, and their comparison with the microscopic calculations is 
discussed. In Sec. 4 we construct a microscopically based density functional
and we study the relevance of the gradient terms in determining
the surface properties of different nuclei.
Sec. 5 is devoted to the possible effects of the pairing
correlations on the equation of state of very asymmetric nuclear matter
at low density. Conclusions are drawn in Sec. 6.
\par\noindent
{\bf 2. The BHF approach to low density asymmetric matter}
\par
Let us consider 
homogeneous nuclear matter at low enough density so that two-body
 correlations
dominate. Of course, at very low density, clustering phenomena can occur, 
like deuteron, triton, helium and alpha particle formation. 
Therefore, we are going to consider densities which are low with respect 
to saturation, but still larger than the ones for the onset of clustering.
Typically, we are considering densities between about one tenth to one
half of the saturation density. In this density region,
microscopic calculations based on Bethe-Brueckner approach are expected 
to be quite accurate, and they will be, therefore, the starting point
for our analysis. The Bethe-Brueckner theory is well documented in
the literature ( see e.g. ref. \cite{book} ) and we will not repeat any details
here. We simply want to mention that we will use the continuous choice
of the single particle spectrum, as described in ref. \cite{3h}, since
it was demonstrated there that the convergence properties are better than
with the gap choice. Three bare forces, the Argonne v$_{14}$ \cite{v14},
the Argonne v$_{18}$ \cite{v18} and the Paris force \cite{paris} will be used. 
As it is well known, these two-body forces have to be supplemented 
by three-body forces in order
to obtain a saturation point close to the empirical one. The semi-empirical
force of ref. \cite{urb1,urb2} is used, with parameters that turn out to be
the same for all the considered two-body forces. It has to be stressed that 
the relevance of the three-body force is restricted to densities around 
and above saturation. In particular, the three-body contribution around
the density $\rho = 0.1$ fm$^{-3}$ is less than $0.2$ MeV and, therefore, it
plays no role in our considerations. The introduction of three-body force
is demanded here only to the aim of
constructing a realistic energy functional,
as it will be discussed in the sequel. Similar considerations apply 
to neutron EoS.
\par
In Fig. 1 we show the results for the energy per particle as a function
of density for symmetric nuclear matter and for pure neutron matter.
As a reference we take the Argonne v$_{18}$ interaction, for which
we did systematic calculations with a small density grid. The corresponding
curves (small circles) have been obtained by a polynomial fitting
to the calculated points. Its explicit form is reported in the Appendix. 
This gives the detailed trend of the 
neutron matter and symmetric nuclear matter equation of state at low
density. At density lower than 0.02 fm$^{-3}$ the equations of state have
been extrapolated imposing that the polynomial crosses the origin.
\par 
We have then calculated few representative points for the 
Argonne v$_{14}$ and Paris potentials.
We see that the results  of the three forces are practically identical in 
this density region, both in the case of symmetric nuclear matter
and pure neutron matter. 
Deviations are known to be present
\cite{A&A} among the different forces at saturation density and above.
This close agreement at low density shows that the equation of state
is determined mainly by the nucleon-nucleon phase shifts, namely by
the on-shell G-matrix.\par
Another way to see this
agreement among the different forces is shown in Fig. 2, where
we display the symmetry energy as a function of density,
a quantity which should be relevant for the properties of nuclear
matter at low densities, in particular in the surface region of nuclei.
Here the symmetry energy is just the difference between the neutron
and symmetric matter EoS (at the same density).
In view of the agreement among the different forces, one can conclude that
the equation of state of symmetric nuclear matter and pure neutron matter
at low density is well determined from microscopic calculations and
they can be considered as established. \par
These results can be used as a reference for more phenomenological
approaches, which are expected to reproduce the microscopic equation of state
if they have to be applied in the low density region. 
\par
In principle calculations can be extended to asymmetric nuclear matter with
an arbitrary value of the asymmetry parameter. However, it is well known that
the dependence of the 
equation of state on the asymmetry parameter $\beta = (\rho_n - \rho_p)/\rho$
can be approximated by 
a quadratic form with very good accuracy \cite{asy},
i.e. the energy per particle $ e = E/A$ at a given total
density $\rho = \rho_n + \rho_p$ and asymmetry $\beta$ can be written
as $e = e_n\cdot \beta^2 + (1 - \beta^2)\cdot e_s$, being $e_n$ and $e_s$
the energy per particle for neutron and symmetric matter, respectively,
at the same density. 
Once the symmetric matter and
pure neutron matter equations of state are determined, it is then possible
to calculate the equation of state for a generic asymmetry.
In Fig. 3 we show a quantitative interpolation for various asymmetries
of the energy per particle for the Argonne v$_{14}$ case. 
The striking point is that for
the asymmetry parameter $\beta \approx 0.75$ the Equation of State
is practically flat with $E/A \sim -400$ keV in the density range considered. 
No such behaviour is found for the phenomenological forces.
 The qualitative explanation for this behaviour of the microscopic
calculation can be as follows. The average kinetic energy per particle
increases as $\rho^{2\over 3}$. If one neglects the density and momentum
dependence of the G-matrix, the potential part at low density increases
(in absolute value) linearly. However the Pauli operator quenches 
the scattering processes, and therefore the absolute value of
the G-matrix is slowly decreasing with density. This means that
the potential energy has an increasing trend that is 
slightly less rapid than linear. It is
conceivable that for a suitable asymmetry the trend of the
potential energy can be mimicked by a density dependence close to
$\rho^{2\over 3}$, which can compensate the increase of the
kinetic energy in a relatively wide density range.

\par\noindent
{\bf 3. Phenomenological forces}
\par
Results for the different phenomenological forces are also reported
in Fig. 1 and Fig. 2. We have considered two Skyrme forces, Sly4
\cite{sly4} and  SkM* \cite{skm} and the finite range Gogny force 
\cite{gog}. They are among the most used ones in nuclear structure
calculations. In particular, the Sly4 force has been adjusted also
in order to reproduce the microscopic results of ref. \cite{Ficks}
for pure neutron matter at high density.
One can see that they are quite close to each other in the
low density symmetric nuclear matter case, with the Gogny force slightly
less binding. The situation is quite different in the pure neutron matter 
case, where a large spread of values for the EoS is apparent. This could be
of course expected, since the Skyrme and Gogny forces are mainly adjusted
to reproduce properties of finite nuclei, where the asymmetry is quite
small. These discrepancies reflect into the symmetry energy displayed 
in Fig. 2. It has to be stressed that at saturation the symmetry energy 
trend for the different forces shows a substantial agreement. Indeed,
at $\rho = 0.16$ fm$^{-3}$ for Gogny force $a_y = 31.62$ MeV,
for SKM* $a_y = 31.35$ MeV, while for the Sly4 force $a_y = 32.72$ MeV.
Thus the trend at saturation is not preserved at lower densities, 
i.e. the value of the symmetry energy at saturation does not fix the
behaviour of the symmetry energy at low density.\par
As already mentioned, for microscopic calculations the trend of the
symmetry energy at low density can be considered independent
of a realistic NN interaction and therefore we think that these results
provide accurate values for the nuclear matter symmetry energy
at low density.
\par
Comparison of the results with the microscopic calculations indicate
that additional constraints on the phenomenological forces can be
introduced if pure neutron matter has to be well described. These constraints
can be of great relevance for the description of the surface region of
exotic nuclei with a large neutron excess. 
The slope of the neutron matter EoS, taken at a reference
value of the density, characterizes the behaviour
of the EoS in the low density region, in particular the amount of binding 
for a given EoS. \par
A comparison similar to the one of Fig. 1,
between microscopic calculations and
phenomenological ones for pure neutron matter only, can be found in ref.
\cite{peth}. The microscopic results for pure neutron matter, both from
variational and Dirac-Brueckner, show a close agreement with the ones
reported in Fig. 1. 
\par
In neutron stars different spatial configurations of almost pure neutron 
matter have been conjectured to occur \cite{peth} just below the solid crust.
These configurations are characterized by non-uniform density profiles,
like rods, ``lasagne", and so on. In particular, neutron bubbles should be
quite sensitive to the detailed trend of the pure neutron matter EoS in the 
considered density range, and therefore an accurate tuning of the 
phenomenological forces, aimed to reproduce the microscopic neutron EoS, 
is probably appropriate.

\par\noindent
{\bf 4. From nuclear matter to finite nuclei}
\par
In order to study the possible link between low density neutron and nuclear
matter EoS and the surface properties of nuclei, we have constructed
a simplified energy functional based on the microscopic calculations
of Sec. 2. This will enable also a more stringent comparison with the
finite nuclei calculations based on Skyrme forces.\par
In the BHF approach the nuclear matter energy density is just the sum of
the free kinetic energy density and the two-body correlation energy
\beq
\barr{rl}
{E\over V} \, & =\, {1\over V}\, \sum_k {{\hbar^2 k^2}\over{2M}} n(k)
 + {1\over {2V}} \sum_{k,k'} \bra{k k'} G \ket{k k'} n(k) n(k') \\
    &            \\
   &=\, \tau (\rho) \, +\, U (\rho) \\    
\ear
\eeq
\noindent
where $n(k) = \Theta (k_F - k) $ is the free Fermion gas occupation number,
$M$ the nucleon mass, $G$ the Brueckner $G$ - matrix, and the variables
$k, k'$ include momentum, spin and isospin quantum numbers. Both kinetic
energy term and correlation energy term are well defined functions of the
density $\rho$. This sum will be identified with the bulk part of the
energy functional. Notice that no effective mass is here explicitly
introduced, at variance with the Skyrme functionals we are considering
(Sec. 3). All the correlation effects are embodied in the interaction term
$U(\rho)$. \par
The full functional appropriate to finite nuclei can be obtained by keeping 
the same correlation energy $U (\rho)$  and including a possible gradient
term of the form $ \eta (\nabla \rho )^2 $. Of course, this term cannot 
be obtained from nuclear matter calculations, and the parameter $\eta$ 
has to be considered as a phenomenological one. 
This energy density functional could be understood within the Kohn-Sham 
(KS)
theory \cite{KS} where the finite range Hartree term has been expanded in
terms of distributions \cite{DSV} providing the gradient term and where the
bulk BHF energy density basically corresponds to the exchange-correlation
part.
In a KS approach based on this functional the hamiltonian density will be
\beq
{\cal H} (\rho) \, =\, {\hbar^2\over 2M} \sum_i (\nabla \psi_i)^2
 \, +\, \left[ U (\rho) + \eta (\nabla \rho )^2 \right] 
\label{ende}
\eeq
\noindent
where $\psi_i $ are the single particle orbital wave functions, 
$\rho (x) = \sum_i \vert \psi_i \vert^2 $ . 

 The wave functions 
$\psi_i$ are determined by minimizing the total energy with the constraint
of particle number conservation. As it is well known, this gives the
KS equations for $\psi_i$ and the total energy of the nucleus
\beq
  E \, =\, \int d^3 x {\cal H} (\rho(x))
\eeq

 To properly describe finite nuclei, the asymmetry and Coulomb energies as 
well the spin-orbit term have to be included. For the density-dependent 
symmetry energy we take the difference between neutron and symmetric matter 
energy densities locally. Also the asymmetry parameter $\beta=(\rho_n - 
\rho_p)/\rho$ is calculated at each point. For asymmetric nuclei we will keep, 
for simplicity, the scalar gradient term of Eq. (\ref{ende}) and neglect the 
possible isovector contributions, which have a small influence and which can 
be 
included in the coefficient of the isoscalar term. The Coulomb energy is 
obtained from the proton density and its exchange term is evaluated in the 
Slater approximation.
 The spin-orbit contribution is equal to the one used in the Skyrme or 
Gogny forces. With this complete functional, which can be included 
within the quasi-local density functional theory \cite{STV} due to the 
spin-orbit term, an explicit link is established between the microscopic   
nuclear and neutron matter EoS and finite nuclei. A similar procedure was 
followed in ref. \cite{xavfaess} for a relativistic extended Thomas-Fermi
calculation of finite nuclei based on a Dirac-Brueckner-Hartree-Fock EoS.

With this functional we performed Kohn-Sham calculations for the magic nuclei
$^{16}$O, $^{40}$Ca, $^{48}$Ca, $^{90}$Zr and $^{208}$Pb. The free parameters 
for our functional,i.e. the parameter $\eta$ and the strength of the 
spin-orbit force $W_0$, are fitted to obtain a good reproduction of the 
experimental binding energy, charge radius and spin-orbit splittings of some 
levels near the Fermi surface of the previously mentioned magic nuclei.

\par At the practical level, the correlation energy
density $U (\rho)$ of the $v_{18}$ EoS was fitted with a simple polynomial
form, to be used in the actual KS calculations, as already mentioned 
in Sec. 2. For densities close to the
saturation one we allowed some small deviations of few hundreds keV from the
calculated energy values, so that the saturation point turns out very close to
the phenomenological one. This deviations are within the numerical accuracy of
the microscopic calculations and are introduced to make the functional as
realistic as possible. The main region of interest for the nuclear surface is
of course the low density region, where the microscopic calculation is 
accurately reproduced.  We got for the saturation density
$\rho_0 = 0.155$ fm$^{-3}$ and the corresponding energy $E / A \, = \, -15.4 $
MeV. The incompressibility and symmetry energy at saturation are 
$K = 247$ MeV and $a_y = 31.34$ MeV, respectively. 
\par 
With this functional we performed KS calculations for a set of
nuclei and varied the parameter $\eta$ in order to get a good reproduction
of the data throughout the periodic table. As a reference we took the value
from the SkM$^*$ force, $ \eta \, =\, -68$ MeV fm$^{5}$. The optimal value
turns out to be $ \eta \, \approx \, -53$. The spin-orbit strength is fixed 
to 130 MeV fm$^5$. Some results are reported in Table
1, in comparison with SkM$^*$, which gives excellent reproduction of the
phenomenological data. Of course, this microscopically based functional cannot
compete with more phenomenological ones, like the SkM$^*$ functional, since
in this case only the parameters $ \eta $ and $W_0$ are adjusted. However, 
this procedure allows to obtain a realistic enough functional as basis for our 
considerations on
the link between surface properties and nuclear matter EoS.
\par 
The results reported in Table 2 for the neutron $r_n$ and proton $r_p$ root 
mean square radii show the possible influence of the gradient 
term on the
value of its difference $ r_n \, -\, r_p $. One can see that this quantity is,
to a large extent, insensitive to the precise value of $ \eta $, and even for
$ \eta \, =\, 0$ its value is not dramatically changed. In other words, the
value of $ r_n \, -\, r_p $ is mainly determined by the bulk part of the
functional. This result provides an explanation, or at least
a justification, of the existing link between
the value of $ r_n \, -\, r_p $ and nuclear matter EoS. 
\par Also the absolute
values of the neutron and proton radii seem to be only marginally affected by
the value of $\eta$. We have also checked that a possible isovector
gradient term (i.e. the gradient of the 
difference between neutron and proton densities) does not affect 
appreciably the results on radii.
On the contrary, as expected, the binding energy is quite
sensitive to the strength of the gradient term, which has a large effect on
the surface energy as it can be seen from Table 2. An estimate of the 
mass formula surface energy coefficient $a_s$ can be obtained by
subtracting from the total energy the bulk energy in asymmetric nuclear 
matter as well as the Coulomb and  spin-orbit contributions, which can be all 
easily extracted from the KS calculations. This coefficient estimated for 
several values of the $\eta$ parameter are also reported in Table 2.
In the case of $^{208}Pb$, one indeed realizes that the surface energy 
reduces by about
a factor 1.7 going from the value $ \eta \, =\, -53$ to $ \eta \, 
=\, 0$ ,  pointing out the key role of the gradient term in the energy 
density to properly describe the nuclear surface.
\par Following refs. \cite{brown,brown2,xavier} we consider the possible
correlation between the difference $ r_n \, -\, r_p $ and the nuclear matter
EoS. Instead of considering the slope of the neutron EoS at the density $ \rho
\, =\, 0.1 $ fm$^{-3}$, we prefer to consider the slope
 $ \chi $ of the symmetry energy $ a_y$ ( at the same density ),
i.e. $ \chi = d a_y / d\rho$. 
As shown in Fig. 4 
the approximate linear correlation, found in refs. \cite{brown,brown2,xavier},
is also valid with this different choice for the EoS parameter. This is 
probably due to the fact that the symmetric matter slope at the reference 
density is quite similar for all forces. The set of relativistic mean field 
functionals clearly cluster on the upper - right part of Fig. 4, while
most of the Gogny and Skyrme forces are concentrated in the lower - left 
part. The point ( the star in Fig. 4 ) corresponding to the 
microscopic functional, constructed
in the present work, lies within the Gogny and Skyrme region.
This indicates that the small discrepancies between the microscopic and Skyrme
forces EoS, as displayed in Fig. 1 and Fig. 2, do not modify appreciably
the considered linear relationship. An accurate measurement of the 
neutron radius in $^{208}$Pb, as projected through parity violating electron 
scattering at Jefferson Laboratory should allow to obtain a value of $r_n - 
r_p$ precise enough to discriminate between different sets of EoS,
 i.e. the one based on relativistic mean field and the one
on Skyrme forces or microscopic functionals, respectively. 
The non-relativistic microscopic functional gives a well
defined prediction for the position of the point along the linear plot.    
The uncertainty contained in this analysis, namely the precise value
of the parameter $ \eta $, is well below the discrepancy between the two
sets of functionals.  

\par\noindent
{\bf 5. Superfluid properties}
\par
It is well known that neutron superfluidity is a pronounced function of 
density and is strongest for about one fifth of the saturation density,
where the gap is about 3 MeV \cite{Lomb}. We therefore can expect that for
$\beta \approx 0.75$ where the EoS is completely flat,  the additional 
pairing correlations can form a local pocket in the EoS where the 
superfluid correlations are strongest. We therefore made a calculation
where in addition to the Brueckner approach we included pairing
correlations in the BCS approximation using the bare force in the gap
equation. The procedure is the same as the one used in ref. \cite{pair}.
Many-body correlations can alter substantially the value of the gap
\cite{zet}. However, the issue is still controversial and we prefer to
use the value calculated within the BCS approach with the bare interaction,
keeping in mind that the results have to be taken as qualitative.
In Fig. 5 we show a blow up of the EoS in the region close to zero binding 
with inclusion of the superfluid phase, again for the Argonne v$_{14}$
potential. We see that indeed a broad depression
around $\rho \, =\, \rho_0/5 - \rho_0/8\,$ is formed which is about $200$
KeV deep. This means that nuclear systems with $\beta \approx 0.75$ at very
low density should be stabilized around this range of density values. For 
example such
a scenario may take place in the outer part of the skin of very neutron rich
nuclei where indeed the $\beta$-values become very large and the neutron 
density drops to low values. Of course we are aware that such type of
picture based on the LDA idea may have only qualitative value. 
Nevertheless it may shed light on the results of more quantitative
structure calculations for finite nuclei.

\par\noindent
{\bf 6. Conclusions}
\par
In this paper we investigated within the Brueckner-Hartree-Fock
approach nuclear and neutron matter at low density typically
between one tenth and one half of the saturation value. 
At these densities the various realistic forces employed give practically
identical results, and we may say that neutron and symmetric matter EoS
are well established in this density domain. 
We also give the EoS as a function of
asymmetry with the help of the well founded quadratic interpolation
formula. We are particularly interested in the EoS for large
asymmetry as they may be found in the surface tail
of neutron rich nuclei but to a less extent also in stable nuclei.
\par
One of the main purposes of the paper is to elucidate the microscopic origin of 
the recently discovered strong correlation
between neutron rms radii of heavy  nuclei and the EoS of neutron matter
\cite{brown} and to what extent this correlation persists within a
microscopic scheme other than semi-phenomenological energy functionals.
We first compared our microscopic EoS in the low density
regime with the ones of various modern Skyrme and Gogny type
phenomenological forces. Though there is qualitative agreement in the
overall behaviour of the EoS there exist important local deviations.
Our microscopic results may therefore serve to fine-tune future
effective forces in the low density regime possibly improving
in this way the surface properties of finite nuclei\par
Then we constructed a realistic density functional based on the
microscopic bulk EoS and applied it to a set of symmetric and 
non-symmetric nuclei. The outcome of this analysis can be summarized as
follows. i) Some differences between the microscopic 
and the Skyrme or Gogny EoS appear at low density. In particular, 
for symmetric nuclear matter around $\rho \approx 0.02 $\,\, fm$^{-3}$
the microscopic EoS appears to have more binding than all the considered
phenomenological ones. 
ii) It is possible to construct an energy functional from the
microscopic EoS which is able to reproduce phenomenological data
on nuclei at a level of precision comparable with the best Skyrme or 
Gogny forces. Only two parameters, the strength $\eta$ of the density 
gradient term and $W_0$ of the spin-orbit force, are adjusted. iii)  It is 
found that the difference 
$r_n \, -\, r_p$ between the neutron and proton rms radii is only marginally
sensitive to the parameter $\eta$. This justifies the possibility of a direct 
link between this radii difference and the bulk nuclear matter EoS.
iv) For the microscopic functional, it turns out that the value of 
the radii difference ( in the lead nucleus ) and the slope of 
the symmetry energy in nuclear matter
at the reference density $\rho = 0.1 $fm$^{-3}$ are in agreement with the
linear correlation found previously for the phenomenological effective
forces. The microscopic prediction is close to the Skyrme or Gogny cases,
while it differs substantially from the predictions based on
relativistic mean field approximation. Therefore,
an accurate enough measurement of the neutron radius in $^{208}$Pb
 could give direct
informations on the bulk nuclear matter EoS via the neutron skin 
$r_n - r_p$.\par
We have also considered the nuclear matter EoS at different asymmetry values.
For the strong asymmetry around $\beta \approx 0.75$ we found as a
particular feature that the microscopic EoS is practically constant
over a quite large range of low density values. Switching on neutron pairing
known to have a strong maximum at low density, we find that a local
minimum in the EoS at $\beta \approx 0.75$ develops for densities
of $\rho_0/5$ to $\rho_0/8$. This local pocket, about 200 keV deep, may
have an important bearing for the extra binding of neutrons in
the far tail of neutron rich nuclei and eventually in the formation of
neutron halos.

\par\noindent
{\bf 6 Acknowledgement.}
\par
One of us (X.V). thanks the support 
under grants BFM2002-01868 (DGI, Spain, FEDER) and 2001SGR-00064 (DGR, 
Catalonia).

\vskip 0.3 cm
\par
\noindent
{\bf Appendix}
\par
In this Appendix we give details on the explicit form of the microscopic
energy functional discussed in the paper. The bulk term of the
energy functional is the sum of the free kinetic energy and of the 
correlation energy part $U( \rho )$. The latter, if we put 
$x = \rho / \rho_0$, with $\rho_0 = 0.155$ fm$^{-3}$ (see Sec. 4),
is given by (energy per particle in MeV)
\beq
\setcounter{equation}{1}
 U( \rho ) =  \left\{ \barc{lr}
                 \sum_{n=1}^6 b_n x^n  &   x< 1  \\
                                 &  \\
                 a_1\cdot (x-1) + a_2\cdot (x-1)^2  &  x > 1 
                 \ear \right.  
\eeq                           
\noindent
with the coefficients $b_1 = -116.734211421291 , b_2 = 310.796396961038 ,
b_3 = -584.213036193276 ,  b_4 = 587.659458730963 , b_5 = -290.107854880778 ,
b_6 =  55.5404612395750 $  ; and  $ a_1 =    -14.4391981216314 ,
a_2 = 16.1424105528328 $.
\par
The two forms match at $x = 1$ ($\rho = \rho_0$) up to the second derivative. 
This functional form must be used up to $\rho = 0.2$ fm$^{-3}$, 
which is the interval where the fit has been performed.
\par
For the pure neutron matter case we have a similar expression, just a
polynomial in $x$,  
\beq  
 U( \rho ) = \sum_{n=1}^5 b_n x^n    
\eeq
\noindent
with $b_1 = -40.1046417029240 ,  b_2 = 41.3865460399823 , 
b_3 = -32.1381502519477 ,  b_4 = 15.1308678586138 , b_5  -2.71394288779458 $,
which is valid in the same density interval.
The symmetry energy can be taken as the difference between the two
functionals, including the kinetic energy parts (see the text).

\newpage
\begin{table}
\caption{Binding energy per particle and charge root mean square radius of 
some magic nuclei obtained with several values of the parameter $\eta$ (see 
text). The same values computed with the $SkM^*$ force and the experimental 
ones are also given.} 
\begin{center}   
\begin{tabular}{l c c c c c c c c c c} 
\hline
\small
&  $\eta=-50$ & &  $\eta=-53$ & &  $\eta=-55$ & &  $SkM*$ & &  $Exp$ \\
\hline
 & $E/A$ & $r_{ch}$ & $E/A$ & $r_{ch}$
  & $E/A$ & $r_{ch}$
  & $E/A$ & $r_{ch}$  & $E/A$ & $r_ {ch}$ \\
\hline
\\
$^{16}$O
  & -8.25 & 2.82 & -8.10 & 2.83 & -8.01 & 2.84 & -7.98 & 2.81 & -7.98 & 2.73
\\
\\
$^{40}$Ca
  & -8.63 & 3.52 & -8.52 & 3.54 & -8.45 & 3.55 & -8.53 & 3.52 & -8.55 & 3.49
\\
\\
$^{48}$Ca
  & -8.87 & 3.55 & -8.76 & 3.56 & -8.68 & 3.57 & -8.75 & 3.54 & -8.67 & 3.48
\\
\\
$^{90}$Zr
  & -8.73 & 4.31 & -8.64 & 4.32 & -8.58 & 4.33 & -8.70 & 4.30 & -8.71 & 4.27
\\
\\
$^{208}$Pb
  & -7.79 & 5.53 & -7.73 & 5.54 & -7.69 & 5.55 & -7.87 & 5.51 & -7.87 &
5.50\\
\hline
\end{tabular}
\end{center}
\end{table}

\begin{table}
\caption{Neutron $r_n$ and proton $r_p$ root mean squre radii (in fm) 
, total binding energy $E$ (in MeV) and the estimate of the mass 
formula surface energy 
coefficient $a_s$ (in MeV) of $^{208}$Pb obtained using several values of the 
$\eta$ parameter (in MeV $\cdot$ fm$^5$.} 
\begin{center}
\begin{tabular}{l c c c}
\hline
\small
& $\eta=0$ &  $\eta=-53$ &  $\eta=-68$  
\\
\hline
\\
$r_n$  & 5.454 & 5.648 & 5.696 
\\
$r_p$  & 5.329 & 5.485 & 5.524
\\
$r_n-r_p$  & 0.125 & 0.163 & 0.172
\\
$E$ & -1891.14 & -1608.46 & -1547.64
\\
$a_s$ & 10.55 & 17.66 & 19.31  
\\
\hline
\end{tabular}
\end{center}
\end{table}

\newpage

\mbox{ }

\newpage

\par
\noindent
{\bf Figure captions}
\vskip 0.3 cm
\par\noindent

Fig. 1.- The energy per particle $E/A$ is plotted as a function of the density,
both for pure neutron matter (upper part) and for symmetric nuclear matter 
(lower part). The symbols correspond to the Brueckner calculations with
realistic forces, Argonne v$_{14}$ (diamonds), Argonne v$_{18}$
(small open circles which correspond to the polynomial fit ) and Paris 
potentials (open squares). The lines correspond
to phenomenological nucleon-nucleon forces, the SkM$^*$ (solid line) and
the Sly4 Skyrme forces (short dashed line), and the Gogny force (long dashed
line).

\vskip 0.3 cm
\par\noindent

Fig. 2.- The symmetry energy at low density calculated in the Brueckner 
scheme and with phenomenological nucleon-nucleon forces. Same notation
as in Fig. 1.

\vskip 0.3 cm
\par\noindent

Fig. 3.- Equation of state at different asymmetries in the parabolic
approximation for the symmetry energy and for the Argonne v$_{14}$
potential. The lowest and higher curves correspond to the nuclear
and neutron matter EoS of Fig. 1 respectively. The other lines correspond
to the EoS of asymmetric nuclear matter from asymmetry $\beta = 0.1$
to asymmetry $\beta = 0,9$ in steps of 0.05 . Some values of the asymmetry
are reported as label of the corresponding EoS.

\vskip 0.3 cm
\par\noindent

Fig. 4.- Linear correlations between the neutron and proton radii and the
slope ( Mev fm$^3$ ) of the symmetry energy at the 
density $\rho = 0.1 $fm$^{-3}$. 
Squares from the top correspond to$NL1$ \cite{NL1}, $NL3$ \cite{NL3}, $G1$, 
$G2$ \cite{Furns} and $Z271$ \cite{Pic}. Triangles to Gogny forces $D280$, 
$D300$, $D250$, $D260$ $D1$ and $D1S$ \cite{Blaizot}. Circles correspond to 
the Skyrme forces $SV$ \cite{Giai}, $SII$ \cite{VB}, $SIV$ \cite{Giai}, $SkM$ 
, $SkM^*$ \cite{skm}, $SLy4$, $SLy5$ \cite{sly4}, $T6$ 
\cite{t6}, $SGII$ \cite{sg}, $SI$ \cite{VB}, $SIII$ and $SVI$ \cite{Giai}.
   \vskip 0.3 cm
\par\noindent

Fig. 5.- Equation of state of nuclear matter at asymmetry $\beta = 0.75$
and with the inclusion of pairing energy.

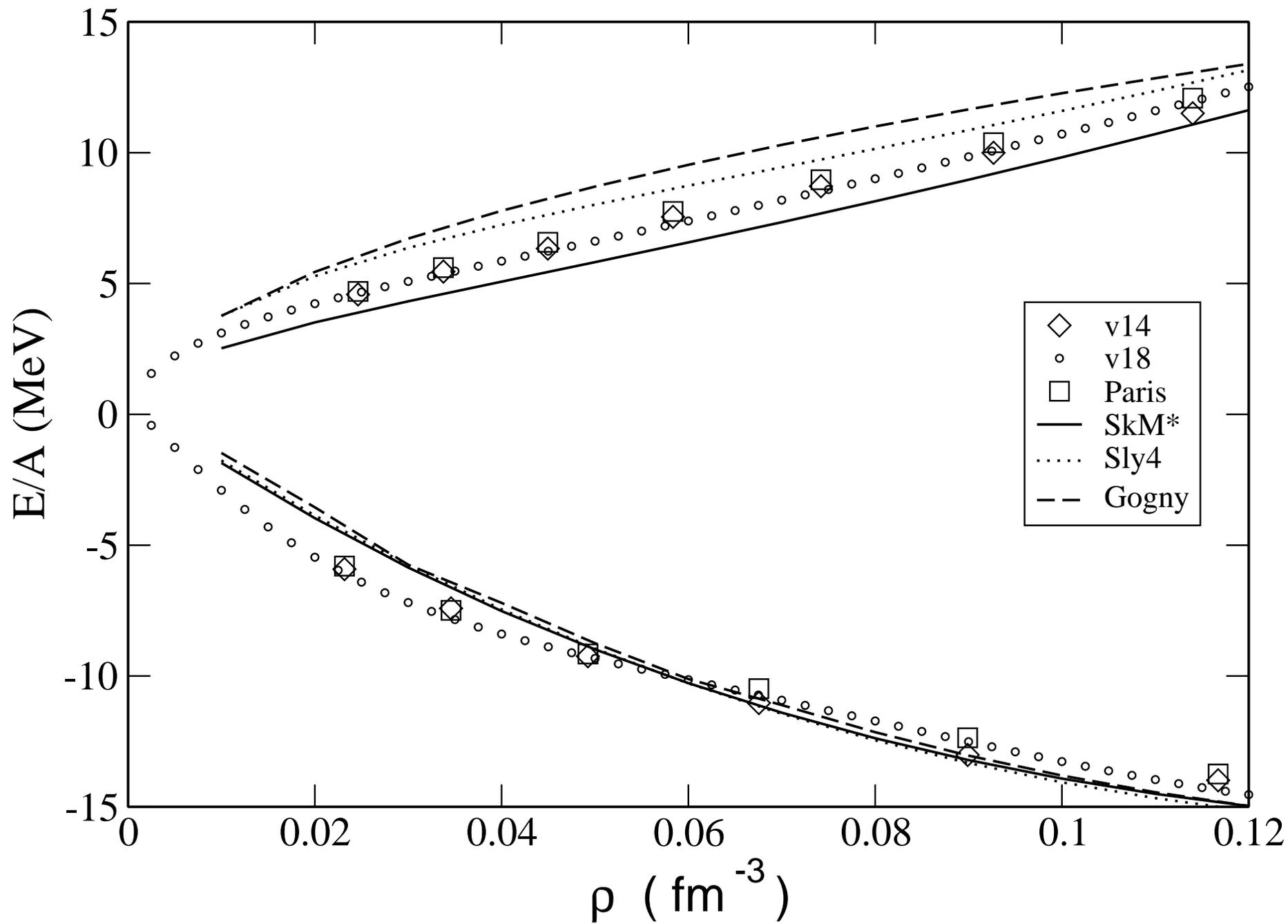
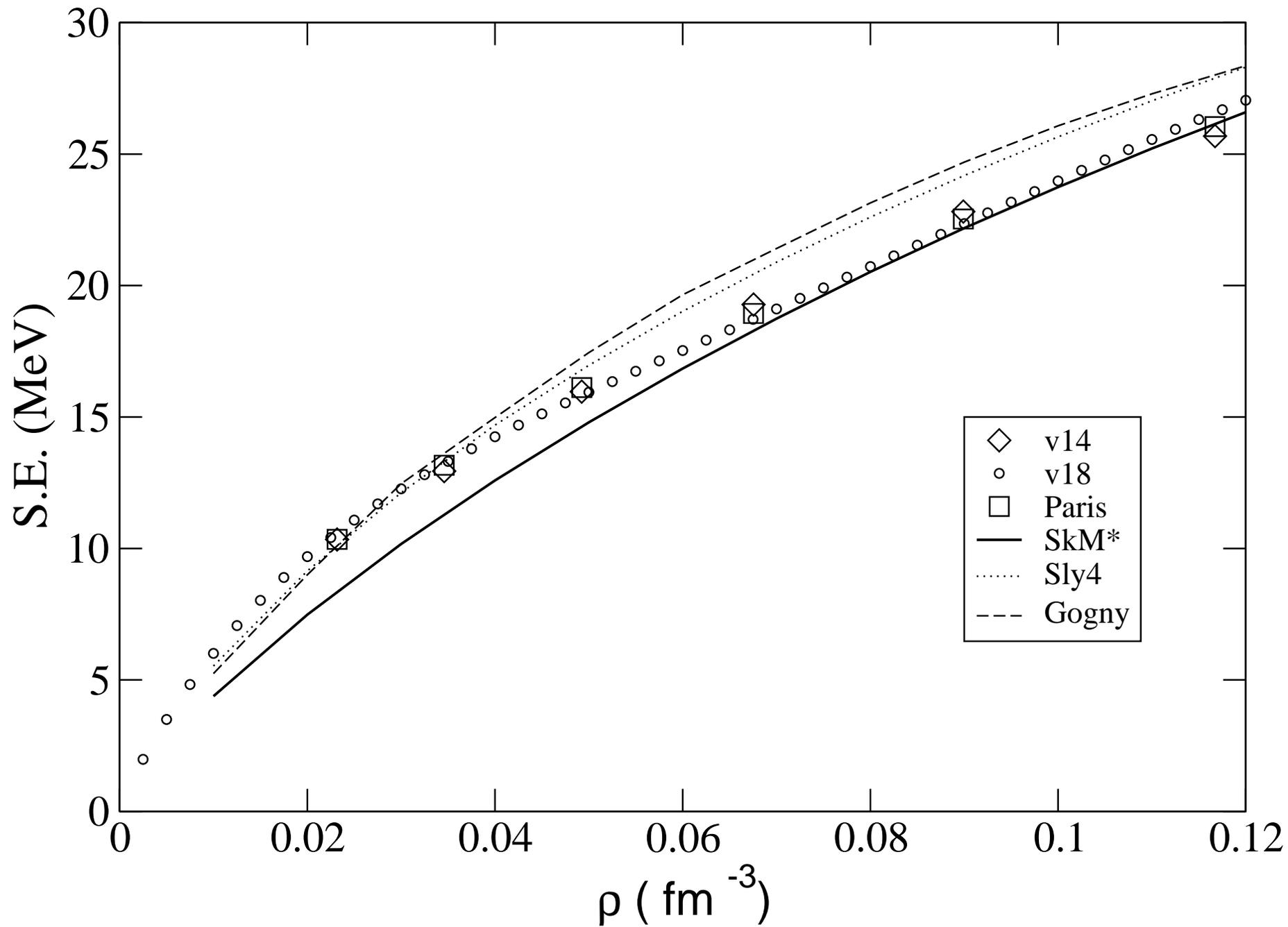
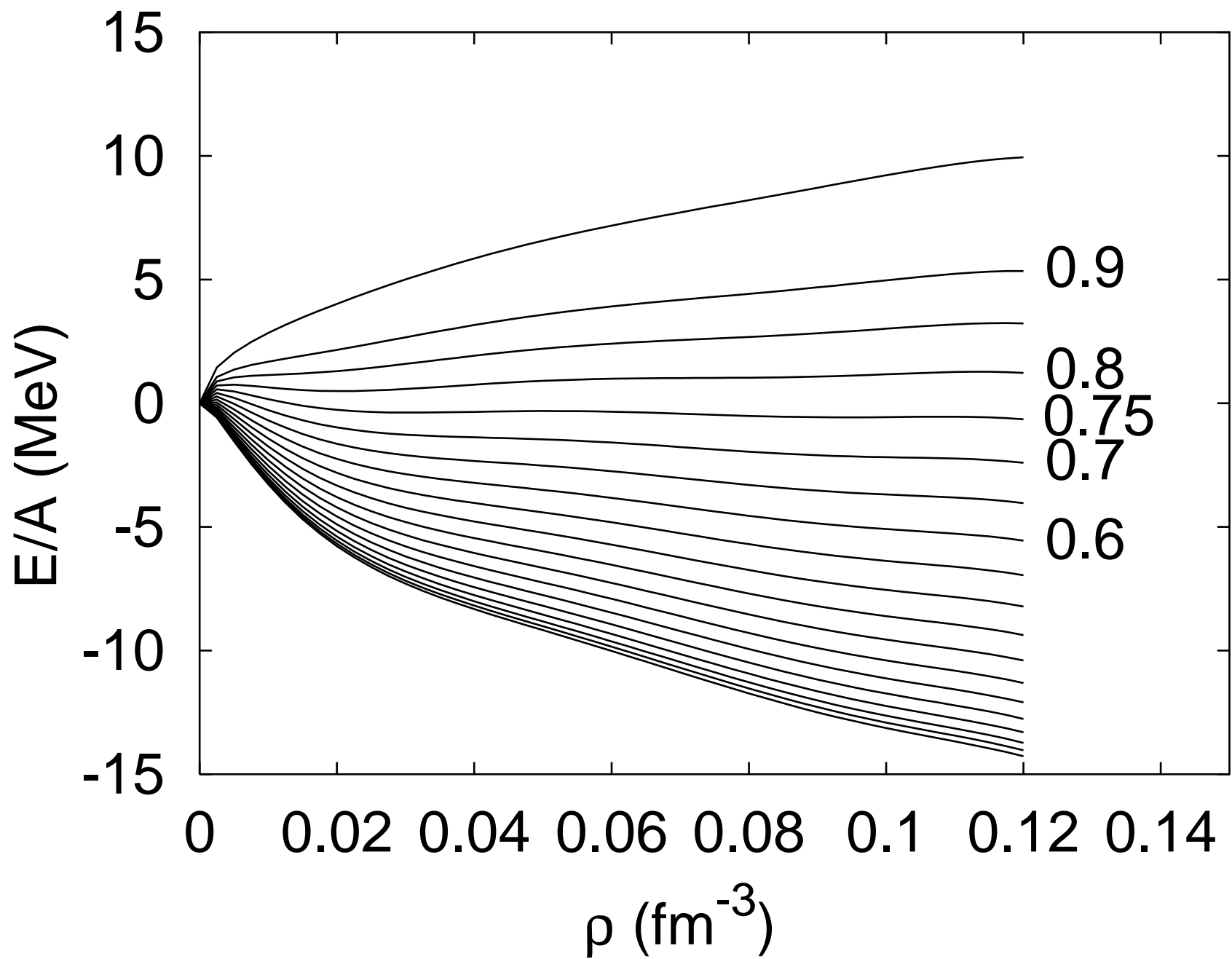
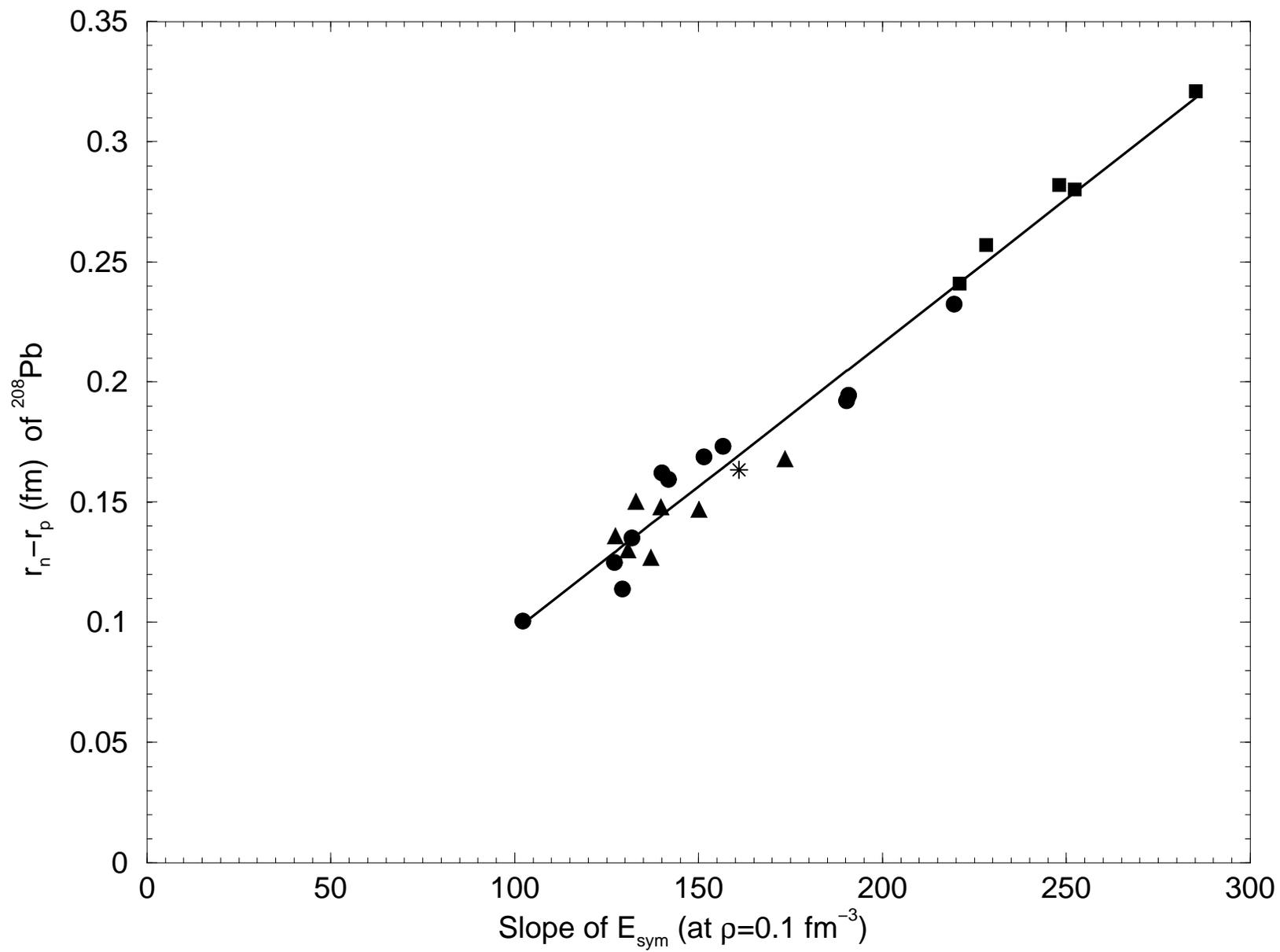
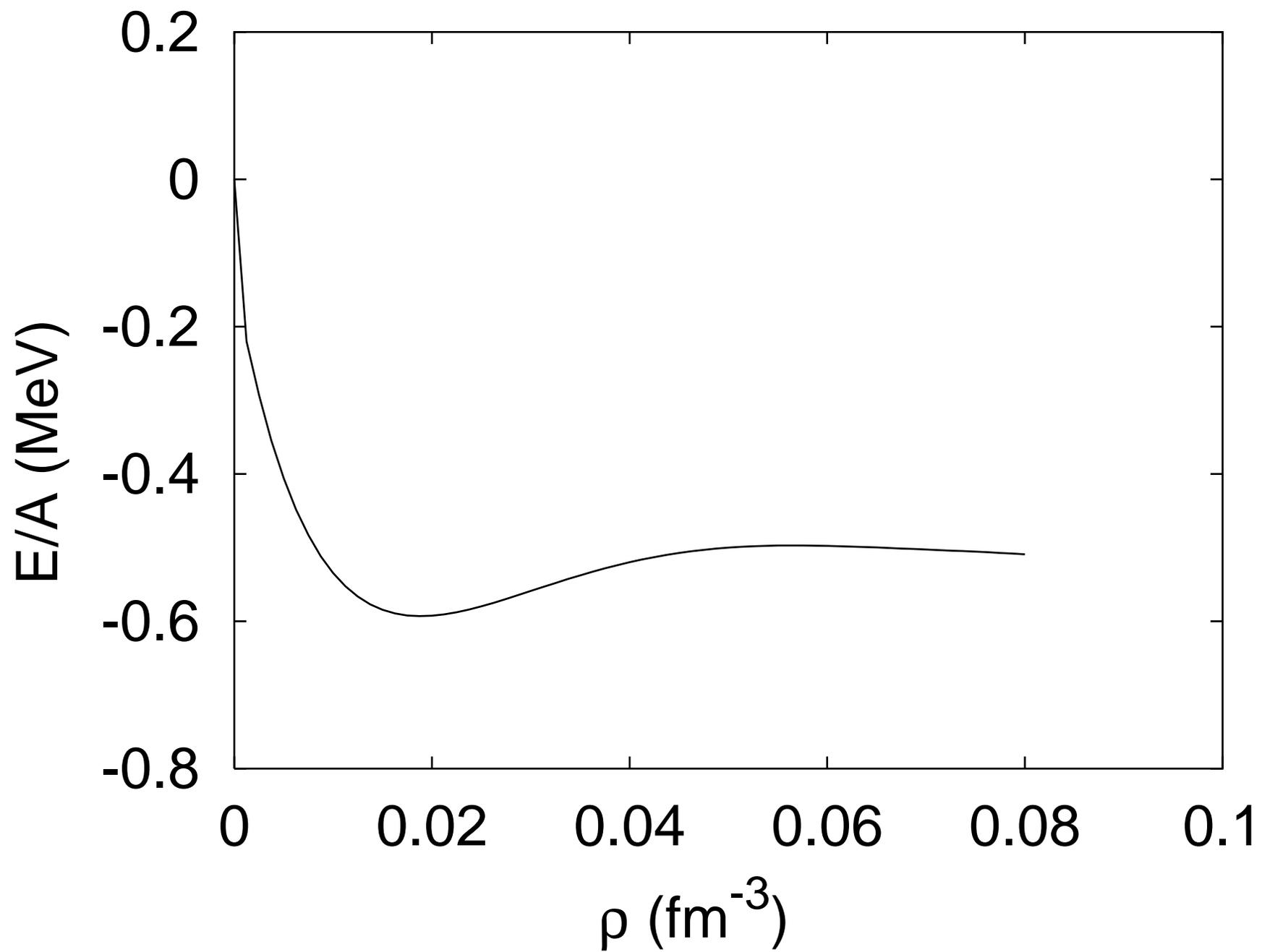
\end{document}